\documentclass[conference]{IEEEtran}
\IEEEoverridecommandlockouts

\usepackage{stfloats}
\usepackage{cite}
\usepackage{amsmath,amssymb,amsfonts}

\usepackage{algorithmic}
\usepackage{graphicx}
\usepackage{subcaption}
\usepackage{textcomp}
\usepackage{xcolor}
\usepackage{amsmath}

\begin{document}
\title{Adaptive Detection of On-Orbit Jamming for Securing GEO Satellite Links}
\IEEEoverridecommandlockouts  
\author{
  \IEEEauthorblockN{Anouar Boumeftah\IEEEauthorrefmark{1}, Olfa Ben Yahia\IEEEauthorrefmark{1}, Jean-François Frigon\IEEEauthorrefmark{1}, Gregory Falco\IEEEauthorrefmark{2}, Gunes~Karabulut~Kurt\IEEEauthorrefmark{1}}
  
  \IEEEauthorblockA{\IEEEauthorrefmark{1}Poly-Grames Research Center, Department of Electrical Engineering, Polytechnique Montréal, Montréal, QC, Canada\\
  \IEEEauthorrefmark{2}Cornell University
Ithaca, NY, US\\
  \textit{\{anouar.boumeftah, olfa.ben-yahia, j-f.frigon, gunes.kurt\}@polymtl.ca, gfalco@cornell.edu }}
  }

\maketitle
\begin{abstract}
This paper introduces a scenario where a maneuverable satellite in geostationary orbit (GEO) conducts on-orbit attacks, targeting communication between a GEO satellite and a ground station, with the ability to switch between stationary and time-variant jamming modes. We propose a machine learning-based detection approach, employing the random forest algorithm with principal component analysis (PCA) to enhance detection accuracy in the stationary model. At the same time, an adaptive threshold-based technique is implemented for the time-variant model to detect dynamic jamming events effectively. Our methodology emphasizes the need for the use of orbital dynamics in integrating physical constraints from satellite dynamics to improve model robustness and detection accuracy. Simulation results highlight the effectiveness of PCA in enhancing the performance of the stationary model, while the adaptive thresholding method achieves high accuracy in detecting jamming in the time-variant scenario. This approach provides a robust solution for mitigating the evolving threats to satellite communication in GEO environments.

\end{abstract}

\begin{IEEEkeywords}
Geostationary orbit (GEO), on-orbit jamming, machine learning in satellite security, satellite communication security, time-variant jamming, satellite interference mitigation.
\end{IEEEkeywords}

\section{Introduction}

Space networks have become vital for both civilian and military operations, playing a key role in global positioning, navigation, and space tracking. The growing reliance on satellite systems raises security concerns, especially for critical infrastructure. Geostationary orbit (GEO) satellites are crucial for continuous coverage and uninterrupted services, making them prime targets for threats such as eavesdropping, spoofing, and jamming \cite{jiang2015security}. Addressing these vulnerabilities is key to ensuring the security of satellite networks (SatNets).

The literature often focuses on active and passive attacks against ground stations, with limited attention to space-based threats. However, real-world incidents, such as the 2022 cyberattack on Viasat's KA-SAT network during the Ukraine conflict, highlight the growing concern about satellite security, as it disrupted communications across Europe \cite{viasat-case-study}. Jamming is a significant threat, particularly in satellite-to-ground and ground-to-satellite links. Most anti-jamming strategies target these links, while satellite-to-satellite links have received less focus. Recent studies, however, are exploring security solutions for space environments. For example, a study on noncooperative jamming in inter-satellite links (ISLs) in mega constellations analyzed jamming power, constellation scale, and inclination, finding that close constellation inclinations maximize jamming efficiency \cite{zhang2024jamming}. Additionally, a location-based authentication framework for secure cislunar space communication has been proposed, using distance verification and adaptive decision criteria to improve link reliability \cite{10592290}. These studies underscore the diverse security challenges in different space missions.

In the current literature, jamming detection methods are broadly categorized into two classes: non-machine learning-based approaches \cite{han2021secure,lichtman2016analysis,taricco2022jamming,planta2024let} and machine learning-based methods  \cite{han2020dynamic,tang2024leo, Aiswarya2022Jamming,9456965,9666744}. For instance, \cite{han2021secure} introduces a secure architecture for space information networks, incorporating beam hopping and relay selection to protect against uplink jamming and downlink eavesdropping. In \cite{lichtman2016analysis}, the feasibility of reactive jamming in satellites is analyzed, and a mitigation strategy is proposed, leveraging coding and interleaving schemes that exploit geometric constraints to reduce the impact of reactive jammers. The work in \cite{taricco2022jamming} addresses camouflage jamming in satellite Internet of Things networks, introducing detection methods that demonstrate the effectiveness of a simple counting technique for identifying jamming attempts while also highlighting system vulnerabilities. Meanwhile, \cite{planta2024let} explores the feasibility of integrating an orbital intrusion prevention system to protect LEO satellites from malicious ground station transmissions, demonstrating the potential of selective jamming through real-life scenario simulations and emphasizing the need for on-orbit security measures.

While non-learning-based methods can be effective in specific scenarios, they become less reliable when dealing with advanced intelligent jammers who can adapt and change their strategies through interaction. As a response, the introduction of artificial intelligence (AI) and machine learning (ML)-based approaches have emerged as robust solutions in detecting and countering such dynamic jamming threats, as these intelligent techniques can adapt to the evolving behavior of attackers, significantly enhancing the resilience and reliability of SatNets \cite{han2020dynamic}. In this context, \cite{tang2024leo} demonstrates the value of ML-based jamming optimization in LEO satellites, leveraging a genetic algorithm for enhanced security against unauthorized broadcasts, while \cite{Aiswarya2022Jamming} underscores the advantages of using random forest, SVM, and neural networks in wireless networks for precise and cost-effective jamming detection. In addition, ML approaches have shown promise in detecting spoofing and jamming attacks for GPS and aerial platforms \cite{9456965,9666744}.

\begin{figure*}[!t]
  \centering
  \includegraphics[width=0.62\textwidth]{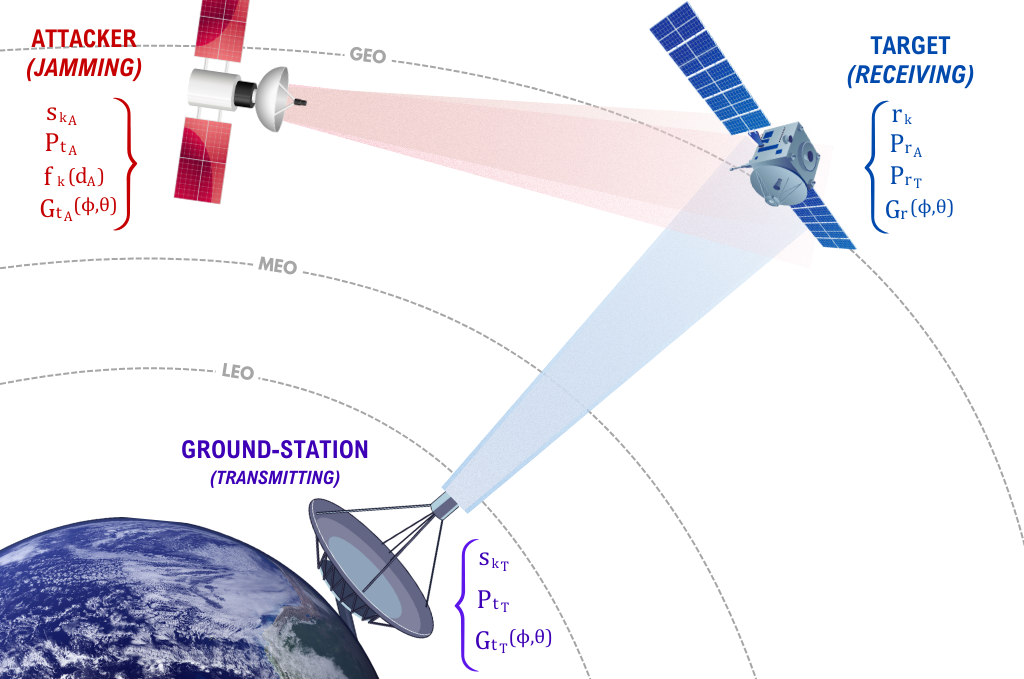}
  \caption{System model illustration: satellite-to-satellite uplink jamming in GEO}
  \label{fig:systemModel}
\end{figure*}

Despite the extensive study of jamming, most existing research primarily addresses ground-based jammers, leaving a significant gap in understanding space-based jamming threats, particularly in GEO orbit. Current studies often assume a stationary jamming model, where the attacker remains fixed, neglecting the complexities of time-variant models, in which the attacker follows a specific trajectory over time. Motivated by these gaps, this paper addresses the emerging threats posed by satellite-to-satellite attacks in GEO, where a GEO attacker jams the uplink communication between a ground station and a GEO satellite, as depicted in Fig. \ref{fig:systemModel}. 

Space-based jammers present unique detection challenges compared to ground-based ones due to their proximity to the target satellite, orbital dynamics, and signal propagation characteristics. Unlike ground-based jammers, which transmit through Earth’s atmosphere, space-based jammers operate in the vacuum of space, avoiding atmospheric attenuation and generating higher power densities. This proximity allows for more localized and directed interference, making detection more difficult. Additionally, space-based jammers often share the same orbital plane as the target, enabling better exploitation of line-of-sight (LOS) communication paths. They can also leverage similar frequency bands used by legitimate systems, further complicating signal differentiation. In contrast, ground-based jammers face obstructions from terrain or environmental factors, limiting their effectiveness and reach.

\noindent Traditional detection methods like directional antenna arrays are inadequate for space-based jamming, as they do not account for orbital motion or space-based geometries. Furthermore, geostationary orbits (GEO) are particularly vulnerable due to the concentration of high-value communication and navigation satellites, making them prime targets for interference. Therefore, this work bridges these gaps by offering novel methods for stationary and time-variant space-based jamming scenarios specifically tailored for GEO environments.
\vspace{2mm}

To address these challenges, we propose a simulation framework and tailored detection methodologies to identify and mitigate these threats. The key contributions of our work are summarized as follows:

\vspace{-2mm}

\begin{itemize} 
\item We identify and contextualize on-orbit jamming threats in GEO, illustrating vulnerabilities arising from proximity-based attacks by maneuverable satellites that can engage in both passive and active jamming. This highlights the necessity of robust on-orbit detection mechanisms.

\item We develop a comprehensive simulation framework for modeling stationary and time-variant jamming attacks in GEO. Using the Systems Tool Kit (STK), we generate realistic satellite trajectories and communication parameters, including signal-to-noise ratio (SNR) and signal-to-jamming-and-noise ratio (SJNR), providing an adaptable tool for future studies in satellite security. 
\item We propose scenario-specific detection approaches, employing Principal Component Analysis (PCA)-augmented Random Forest for stationary jamming scenarios and an adaptive threshold-based method for time-variant attacks. Our methodology incorporates basic orbital dynamics, using satellite dynamics to enhance the accuracy and robustness of jamming detection in GEO environments.

\end{itemize}

\vspace{-2mm}
The remainder of the paper is organized as follows. Section \ref{Sec:case} introduces a case study on anomalous proximity maneuvers by a GEO satellite, highlighting risks associated with proximity-based threats in GEO. Section \ref{Sec:model} provides a comprehensive description of the channel characteristics and the proposed system model. Section \ref{Sec:methodology} details the proposed method for stationary and time-variant jamming scenarios. The numerical results are discussed in Section \ref{Sec:results}. Finally, Section \ref{Sec:concl} concludes the paper.


\section{Case Study: Anomalous Proximity Maneuvers by a Geostationary Satellite}
\label{Sec:case}
In recent years, the security of GEO satellites has faced increased concern, mainly due to the potential threats associated with maneuverable satellites that perform frequent proximity operations. While much of the existing literature focuses on jamming and spoofing in ground-to-satellite, possible attacks involving proximity operations in GEO remain underexplored. 
To address this, we present a case study on a fictional satellite, designated Satellite X, which exhibits highly atypical behavior by repeatedly maneuvering close to other GEO satellites, labeled Satellites A, B, C, and D. This behavior highlights the potential risks of proximity-based interference or surveillance in GEO, emphasizing the importance of developing detection and mitigation strategies for such space-based threats.

\paragraph{Anomalous Maneuvering Patterns}
In this scenario, we assume that Satellite X regularly adjusts its position through intentional orbital maneuvers. Unlike typical GEO satellites, which generally maintain near fixed longitudes, Satellite X periodically maneuvers, pausing near specific locations before initiating another adjustment. This pattern of frequent repositioning deviates notably from standard GEO satellite operations, which aim to maintain a stable orbit with minimal adjustments. Furthermore, Satellite X is observed to position itself near the locations of communication satellites A, B, C, and D.
\begin{figure}[!t]
  \centering
  \includegraphics[width=3.5in]{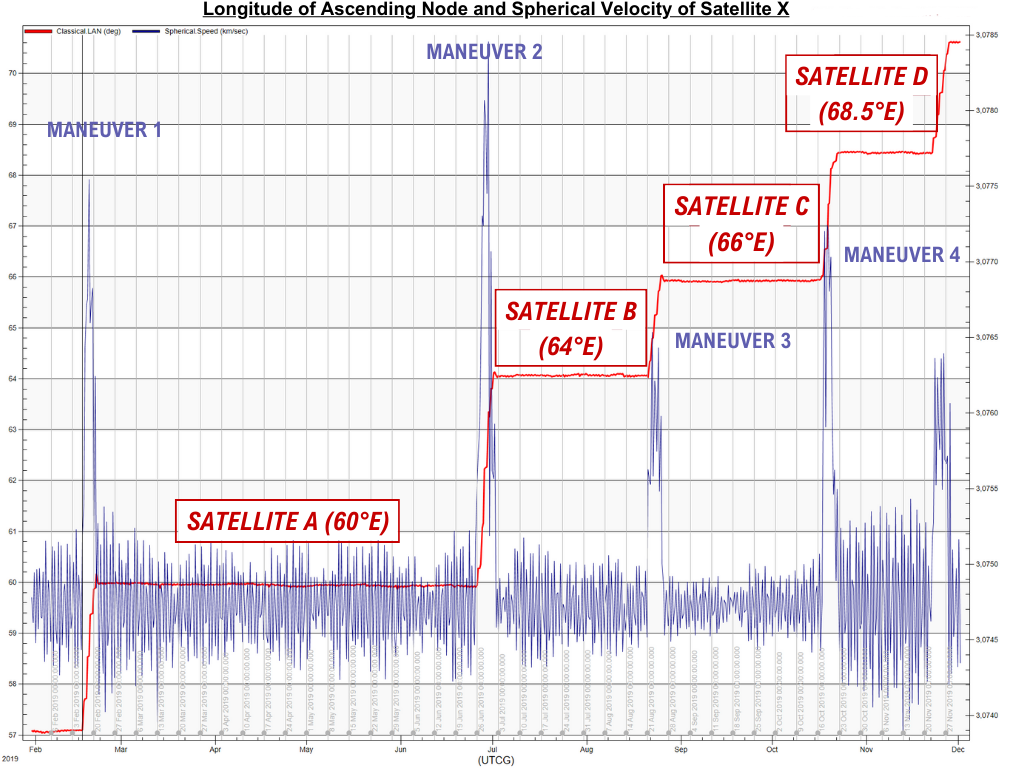}
  \caption{Orbital maneuver patterns of Satellite X: longitude of ascending node and spherical speed over time using STK}
  \label{fig:Orbital maneuver}
\end{figure}
\paragraph{Visualizing the Maneuvers and Orbital Data}

To illustrate this behavior, Fig. \ref{fig:Orbital maneuver} depicts a graph tracking the Longitude of the Ascending Node and spherical speed of Satellite X over time. In this figure, each abrupt change in Longitude of Ascending Node and speed indicates a maneuver, followed by a period where Satellite X remains stationary near the position of one of the communication satellites, frequently aligning with Satellites A, B, C, or D. This pattern of approach-and-hold suggests an intentional strategy of positioning Satellite X near critical communication assets. Such a sequence of maneuvers is highly atypical for a satellite in GEO, where fuel efficiency and orbital stability are primary operational goals. The repeated adjustments imply a strategic intention, potentially increasing the risk of eavesdropping, signal interception, or jamming for the targeted communication satellites.

\paragraph{Challenges and Importance of Dynamic Detection}
The dynamic nature of Satellite X’s maneuvering presents unique interference detection and mitigation challenges. Traditional detection models, often optimized for static interference sources, may fail to accurately identify jamming or interception attempts stemming from a moving source. In this case study, Satellite X’s proximity to Satellites A, B, C, and D and its repeated approach-and-hold maneuvers highlight the need for a time-variant detection model capable of responding to the satellite’s dynamic behavior.

\section{Channel and System Model}
\label{Sec:model}
\subsection{System Model}
As illustrated in Fig. \ref{fig:systemModel}, the system model comprises a legitimate GEO satellite and an attacker operating in the same orbit. The legitimate satellite communicates with a ground station on Earth, while the attacker functions in two distinct modes: passive and active. The attacker transmits jamming signals toward the legitimate satellite and may operate stationary or time-variant. In the stationary approach, the attacker remains fixed at a particular location. In contrast, in the time-variant approach, the attacker adjusts its trajectory over time to maximize jamming impact while minimizing detection risk. Unlike the stationary jamming model commonly assumed in current studies, this dynamic time-variant approach—where the attacker alternates between different positions—introduces significant complexities for detection and mitigation. 

For the uplink, given the significant distance between the ground station and the GEO satellite, coupled with the ground station's limited transmission power, the transmitted data packets are more vulnerable to jamming attacks from an adversary. This risk becomes more prominent when the ground station employs basic or weak encryption. In the downlink, an attacker can intercept the communication due to the broadcast characteristics of satellite channels, posing potential threats to data security and privacy. In the literature, downlink jamming is considered less frequent because it is assumed that a jammer requires a direct LOS to the ground station to generate significant jamming power at the receiver's input \cite{hofmann2018satellite}. Uplink jamming relies heavily on obtaining detailed information about the target signal and requires considerable transmitter power to interfere with the satellite’s radio receivers, including sensors and command systems. Although more challenging to execute, uplink jamming can have far-reaching consequences, potentially disrupting the satellite's service globally for all users \cite{threatsspace}. While uplink jamming in the literature typically assumes a ground-based attacker, our model considers an attacker in GEO, thus representing an on-orbit uplink jamming scenario.

\subsection{Communication Model}
The received signal at the legitimate satellite is modeled 
for each sample \( k \) as: 

\begin{equation}
r_k = \frac{s_{T,k}}{\sqrt{\text{loss}_{T}}} + n_k + \frac{f_k \cdot s_{A,k}}{\sqrt{\text{loss}_{A}}},
\end{equation}
where \( r_k \) represents the \( k \)th signal sample received at the legitimate satellite at time \( t \), \( s_{T,k} \) denotes the \( k \)th modulated signal sample transmitted from the ground station, and \( s_{A,k} \) indicates the \( k \)th transmitted signal sample from the attacker's transmitter. The noise at the \( k \)th sample, \( n_k \), follows a complex normal distribution, denoted as \( n_k \sim \mathcal{CN}(0, \sigma_{\text{noise}}^2) \). The term \( \text{loss} \) refers to the free space path loss (FSPL), and the indicator function \( f_k \) takes the value of \( 1 \) when jamming is present at sample \( k \) and \( 0 \) otherwise.
The average transmitted power of the ground station, \( P_T^{\text{avg}} \), and the jammer, \( P_A^{\text{avg}} \), are determined by the expected values of the squared magnitudes of their transmitted signal samples, \( s_{T,k} \) and \( s_{A,k} \), respectively. These relationships are mathematically represented as follows:
\begin{equation}
P_T^{\text{avg}} = \mathbb{E}\left[\left|s_{T,k}\right|^2\right], \quad P_A^{\text{avg}} = \mathbb{E}\left[\left|s_{A,k}\right|^2\right] .
\end{equation}
Similarly, the received signal strength (RSS) is calculated from the received signal samples \( r_k \) as:
$\text{RSS} = \frac{1}{N} \sum_{k=1}^{N} \left| r_k\right|^2$, where \( N \) is the total number of samples.
{
Finally, we define the SJNR as a key metric to assess the quality of a received signal under noise and intentional jamming. It is given by:
\[
\text{SJNR} = \frac{P_T^{\text{avg}} G_T^T G_R  \frac{1}{\text{FSPL}_\text{Uplink}}}{\sigma_{\text{noise}}^2 + P_A^{\text{avg}} G_T^A G_R  \frac{1}{\text{FSPL}_\text{Attacker}}}
\]

\noindent where \( G_T^T \) and \( G_T^A \) are the gains of the ground station's transmitting antenna (target) and the jammer satellite's antenna (attacker), respectively, \( G_R \) is the gain of the target satellite's receiving antenna, \( \text{FSPL}_\text{Uplink} \) and \( \text{FSPL}_\text{Attacker} \) are the free-space path losses for the ground and attacker paths, respectively. Note that when \( P_A^{\text{avg}} = 0 \), the SJNR reduces to the regular SNR.

}

Periodic jamming events are simulated at specific epochs, labeling data points as ``jammed'' or ``non-jammed.'' All resulting trajectory and communication data are stored in CSV files for subsequent analysis.

\section{Proposed Methodology}
\label{Sec:methodology}
This section outlines the scenario and data generation methodology for the proposed stationary and time-variant models. 
\subsection{Stationary Model}
In this subsection, we describe the orbital dynamics and spatial configuration underlying the stationary simulation based on the system model shown in Fig. \ref{fig:systemModel}. This simulation evaluates the target satellite’s vulnerability to jamming by examining various fixed positions the attacker might occupy within a defined volume of interest (VOI).

The ground station maintains a direct LOS with the target satellite, positioned directly beneath it on Earth’s surface. Due to the orbit's geostationary nature, the target satellite remains stationary relative to the ground station, ensuring a stable communication link. The attacker, also in a GEO position, is a potential threat to this communication link. Possible attacker positions are generated within a predefined VOI surrounding the target satellite to simulate multiple jamming scenarios. The VOI is represented as a spherical region centered on the target satellite, with a radius \( R_{\text{VOI}} \), establishing the spatial domain within which the attacker can maneuver. The spherical constraint for the attacker’s position is defined as:
\begin{equation}
    \| \mathbf{y}_A^{(i)} - \mathbf{y}_T \| \leq R_{\text{VOI}},
    \label{eq:ROI}
\end{equation}
where \( \mathbf{y}_A^{(i)} \) represents the \( i \)-th position of the attacker within the VOI, \( \mathbf{y}_T \) denotes the fixed position of the target satellite in the geocentric coordinate system, and \( R_{\text{VOI}} \) is the VOI radius, set to a value such as 5,000 km to cover a significant critical area around the target satellite. Each attacker position \( \mathbf{y}_A^{(i)} \) within the VOI is sampled uniformly to simulate various jamming scenarios. By generating multiple attacker positions within the VOI, we create a range of locations, each representing a distinct potential jamming scenario. Given that the target and attacker satellites are in GEO, their relative position remains fixed unless the attacker executes deliberate maneuvers. This stationary assumption simplifies the orbital dynamics, allowing the analysis to concentrate on intentional jamming tactics rather than accounting for orbital perturbations.

\subsection{Time-Variant Model}
This subsection introduces a time-variant approach for detecting satellite-to-satellite jamming in scenarios where the attacker is in motion relative to a stationary target satellite in a GEO.
\paragraph{Implementation and expected impact}
Unlike the stationary model, where an attacker's position is fixed and a binary classifier detects threats at time \( t \), the time-variant model adjusts detection thresholds dynamically based on signal variations, enhancing responsiveness to changing conditions. The attacking satellite’s movement along its orbit introduces temporal signal changes, which the model tracks by recalculating adaptive thresholds and monitoring the rate of signal feature changes. This adaptability improves the model’s accuracy in detecting transient jamming across various attack trajectories.

We use the Systems Tool Kit (STK) software and its API to generate satellite trajectories to simulate the time-variant jamming detection. Multiple attacking satellites are simulated, each with randomized orbital parameters such as semi-major axis and inclination relative to a target satellite in GEO. The simulation begins by defining the target’s fixed position above a ground station, then generating and propagating each attacker’s orbit. Access intervals between attacker and target satellites are calculated, and range data (indicating relative distance) is extracted from the AER (Azimuth, Elevation, Range) report. Initially, the study focuses on range data alone, excluding factors including attitude, line of sight, and antenna pointing.

\begin{figure}[!t]
  \centering
  \includegraphics[width=0.9\columnwidth]{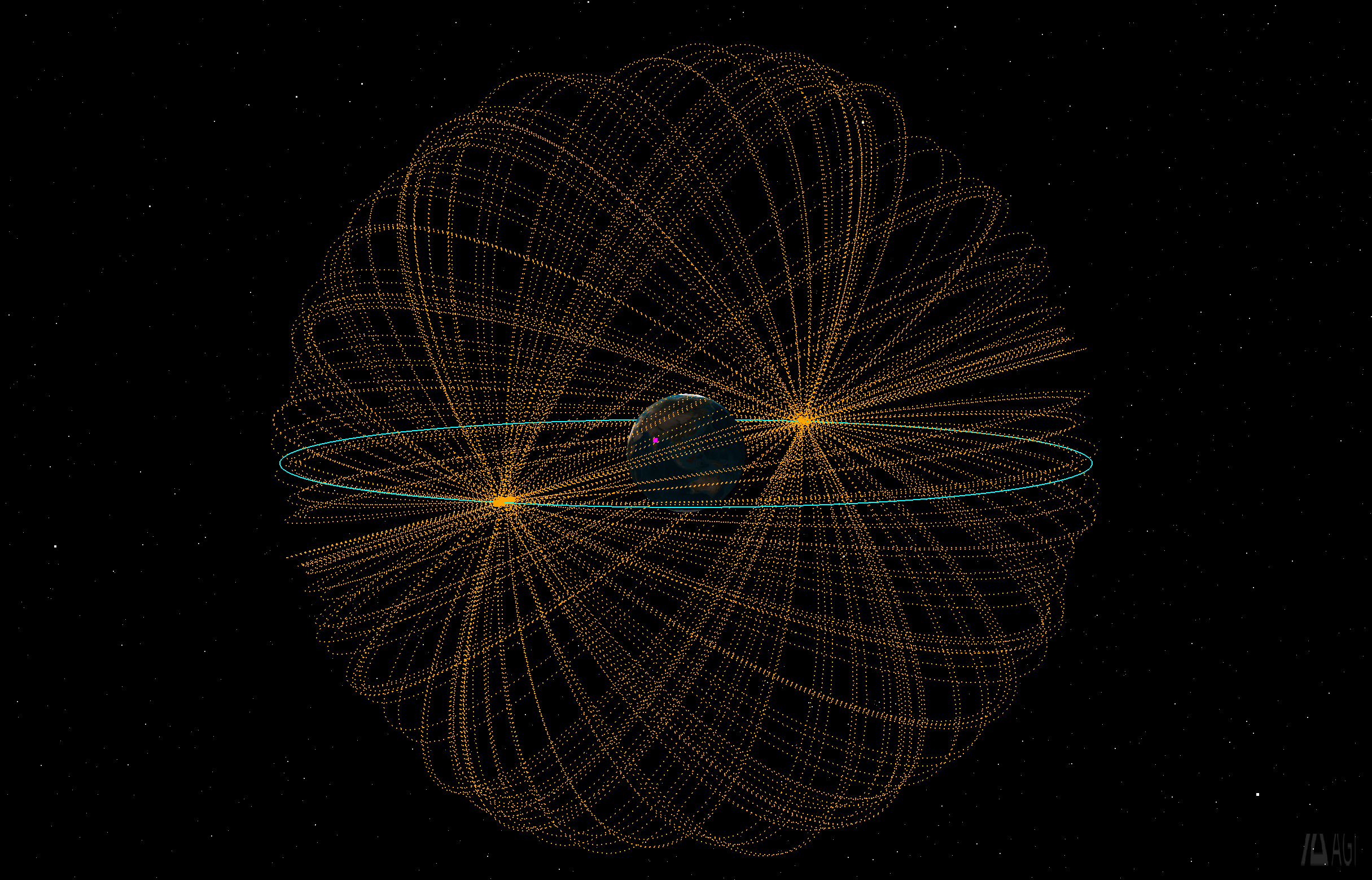}
  \caption{Orbital visualization of the 100 automatically generated trajectories via STK Engine}
  \label{fig:Orbital trajectories}
\end{figure}

\begin{figure}[!t]
  \centering
  \includegraphics[width=0.9\columnwidth]{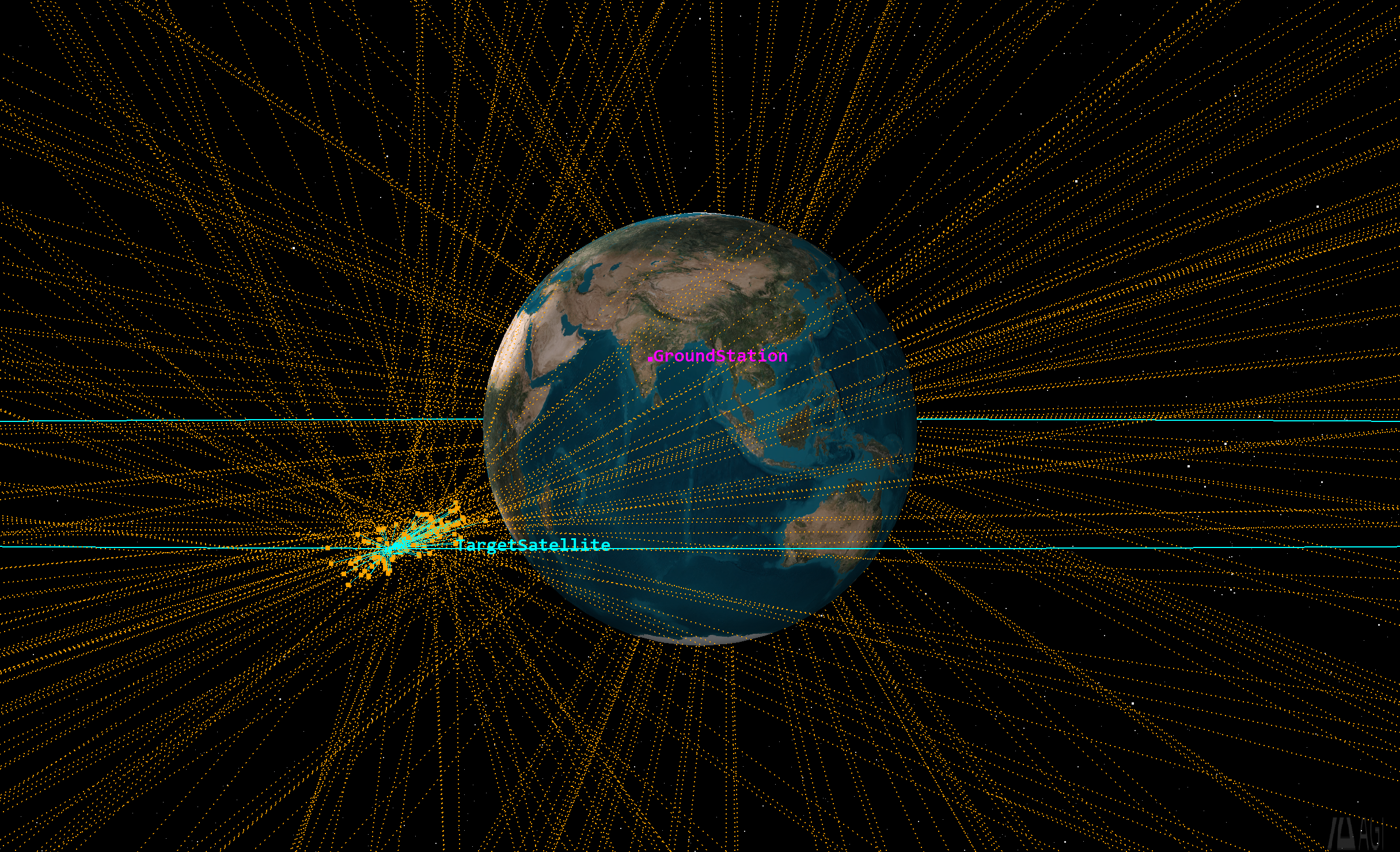}
  \caption{Visualization of the 100 access links of the VOI-constrained attackers generated via STK Engine}
  \label{fig:Orbital positions with access}
\end{figure}

Fig. \ref{fig:Orbital trajectories} shows 100 attacker satellite positions around Earth, highlighting proximity and density relative to the target. Denser attacker groups pose higher jamming risks. Fig. \ref{fig:Orbital positions with access} illustrate direct LOS connections between attacker satellites and the target, revealing potential jamming risks. By analyzing these links and signal metrics, such as SJNR and RSS, we identify the most disruptive attacker positions within the VOI.

In addition, we calculate signal and power parameters at each epoch within the access interval to simulate communication and jamming interactions. This includes modeling the communication link between the ground station and the target satellite and the link between the attacker and the target. This allows the measurement of critical metrics such as received power, SNR, and SJNR. Finally, periodic jamming events are simulated at specific epochs, labeling data points as "jammed" or "non-jammed." All resulting trajectory and communication data are stored in CSV files for subsequent analysis.


\paragraph{Adaptive thresholding approach}
We consider adaptive thresholds and rate-of-change metrics to detect jamming events in the time-variant model, recalculated within a moving window. The key steps of the detection method are outlined as follows:
\begin{enumerate}
    \item Data processing across multiple trajectories: each trajectory file represents a unique relative path the attacking satellite takes concerning the target. We analyze signal features such as SJNR and total RSS for each trajectory to identify potential jamming events.

    \item Dynamic thresholds calculation: adaptive thresholds are determined for each time step \( t \) using a moving window of size \( W \). Within this window, the mean \( \mu \) and standard deviation \( \sigma \) of each feature are computed. Two key parameters, \(\alpha\) and \(\beta\), are used to adjust the sensitivity for thresholding and rate-of-change detection.

\begin{itemize}
    \item \(\alpha\), the threshold multiplier, adjusts the sensitivity of each feature’s threshold by scaling with the standard deviation. A higher \(\alpha\) value decreases sensitivity to fluctuations, while a lower \(\alpha\) increases sensitivity to signal changes.
    
    \item \(\beta\), the rate-of-change threshold is used to detect abrupt deviations in feature values. This parameter sets the threshold for the rate of change in features such as SJNR and RSS between consecutive time steps, marking significant variations as potential jamming indicators.
\end{itemize}
The adaptive thresholds for SJNR and total RSS are calculated as follows:
\begin{align}
    \text{Threshold}_{\text{SJNR}}(k) &= \mu_{\text{SJNR}}(k) - \alpha \cdot \sigma_{\text{SJNR}}(k), \label{eq:threshold_sjnr} \\
    \text{Threshold}_{\text{RSS}}(k) &= \mu_{\text{RSS}}(k) + \alpha \cdot \sigma_{\text{RSS}}(k), \label{eq:threshold_rss}
\end{align}
where \( \mu_{\text{SJNR}}(k) \) and \( \sigma_{\text{SJNR}}(k) \) represent the mean and standard deviation of the respective signal feature over the moving window of size \( W \) at sample index \( k \).

\item Rate of change analysis: to detect abrupt changes in signal characteristics caused by the attacker's movement, the rate of change for SJNR and RSS between consecutive samples \(k-1\) and \(k\) is calculated as 
\(\Delta_{\text{SJNR}}(k) = \left| \text{SJNR}(k) - \text{SJNR}(k-1) \right|\) 
and 
\(\Delta_{\text{RSS}}(k) = \left| \text{RSS}(k) - \text{RSS}(k-1) \right|\), respectively. Significant shifts in these values may indicate jamming activity.

\item Jamming detection criteria: at each sample \( k \), the following detection conditions are evaluated:
\begin{itemize}
  \item Threshold condition: a jamming event is flagged if 
    \(\text{SJNR}(k) < \text{Threshold}_{\text{SJNR}}(k)\) 
    or 
    \(\text{RSS}(k) > \text{Threshold}_{\text{RSS}}(k)\).
    \item Rate-of-Change Condition: a jamming event is flagged if 
    \(\Delta_{\text{SJNR}}(k) > \beta\) 
    or 
    \(\Delta_{\text{RSS}}(k) > \beta\).
\end{itemize}
The communication is labeled as "jammed" if one of the above conditions is satisfied.
    
\end{enumerate}

\section{Numerical Results and Discussion}
\label{Sec:results}

\begin{figure*}[!t]
    \centering
    \begin{subfigure}[b]{0.29\textwidth}
        \centering
        \includegraphics[width=\textwidth]{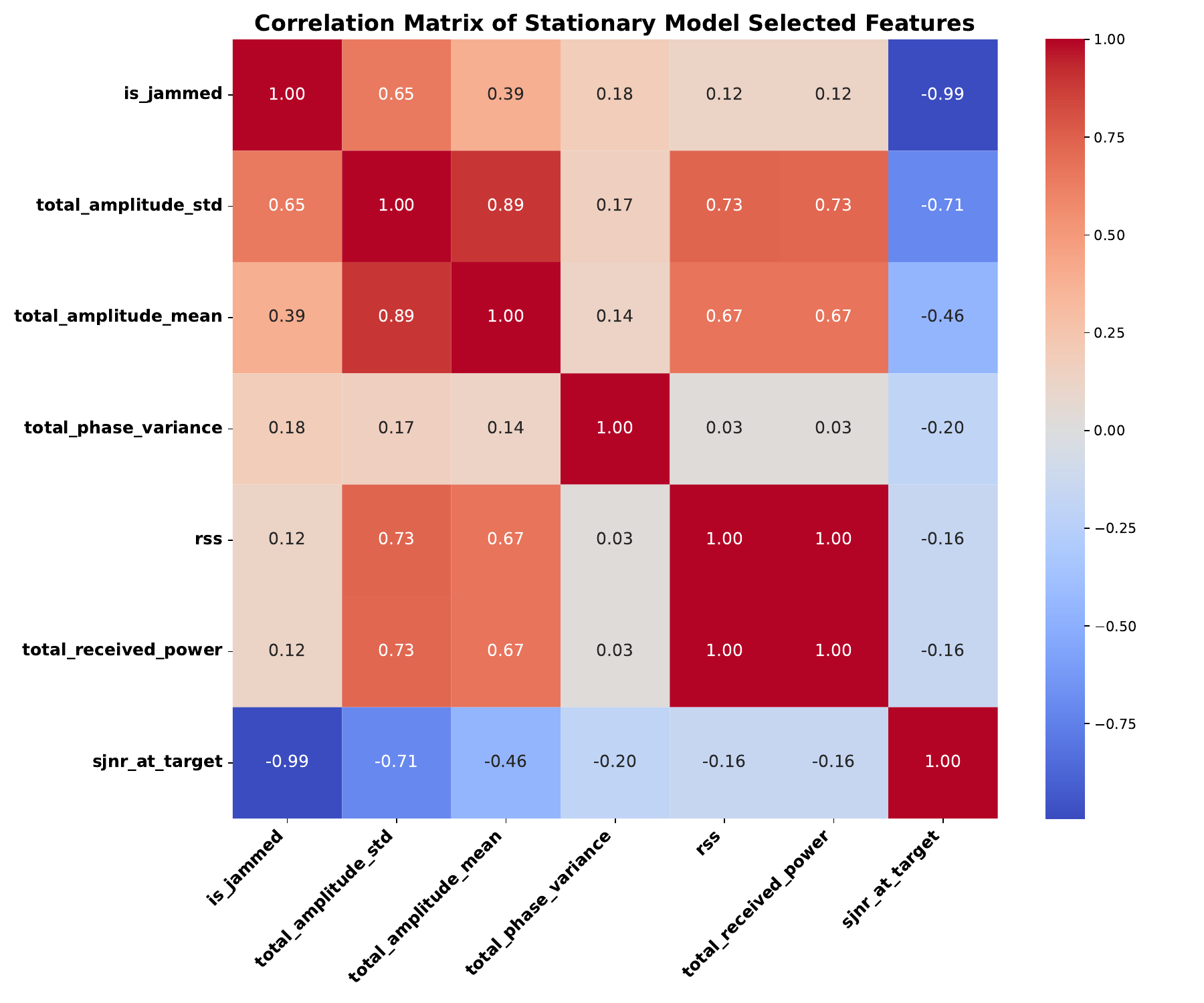}
        \caption{Feature selection on signal metrics using correlation matrix}
        \label{fig:correlation_matrix}
    \end{subfigure}
    \hfill
    \begin{subfigure}[b]{0.29\textwidth}
        \centering
        \includegraphics[width=\textwidth]{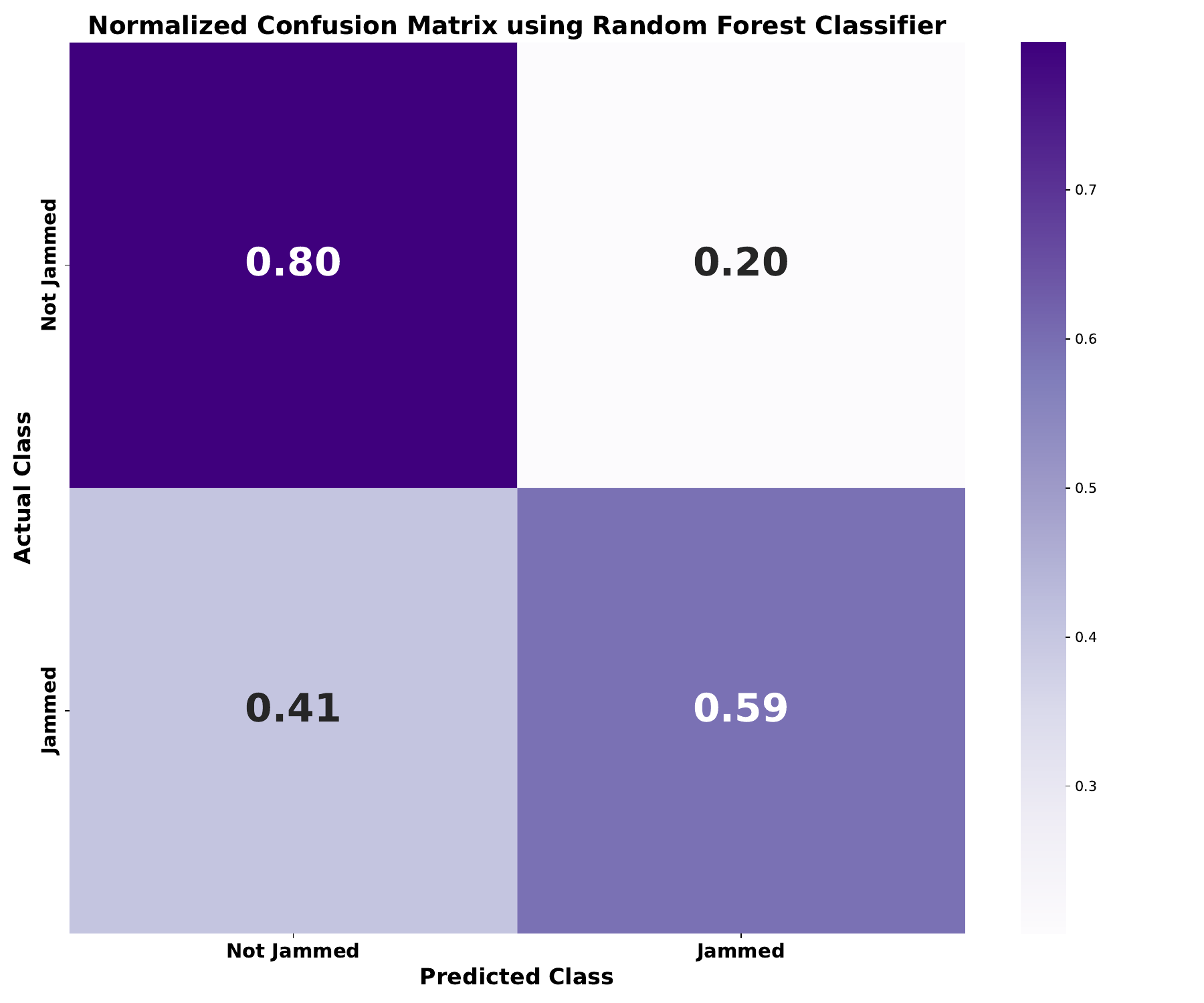}
        \caption{Confusion matrix without PCA}
        \label{fig:confusion_without_pca}
    \end{subfigure}
    \hfill
    \begin{subfigure}[b]{0.29\textwidth}
        \centering
        \includegraphics[width=\textwidth]{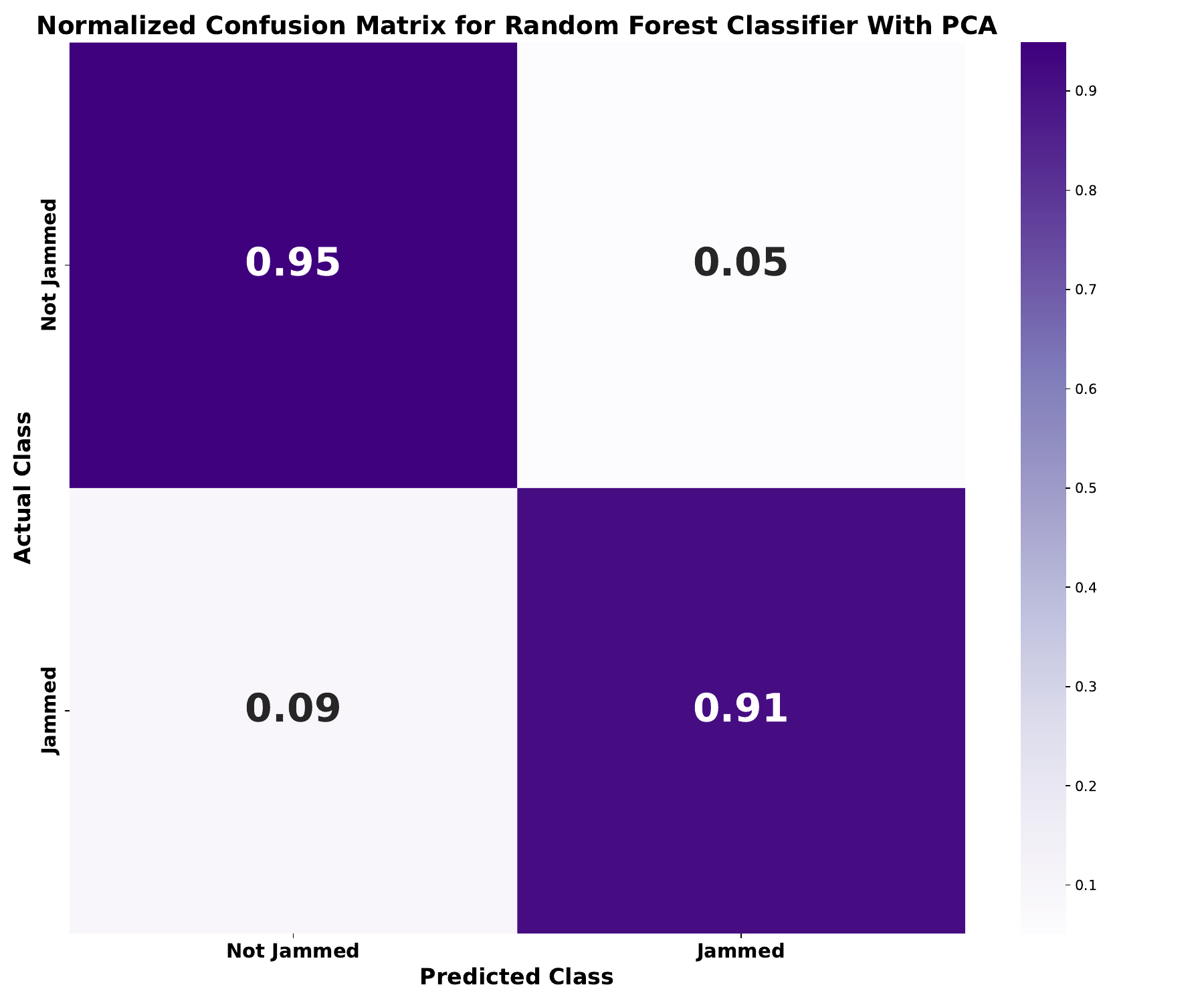}
        \caption{Confusion matrix with PCA}
        \label{fig:confusion_with_pca}
    \end{subfigure}
    \caption{Comparative analysis of feature correlation and performance for the stationary model. (a) shows the correlation matrix for feature selection, indicating relationships and potential redundancies among features. (b) and (c) display confusion matrices for the random forest model without and with PCA, respectively, demonstrating improved classification accuracy with PCA applied.}
    \label{fig:combined_analysis}
\end{figure*}
This section outlines the performance metrics used for the system's evaluation. The carrier frequency is set at 14 GHz with a bandwidth of 1 MHz, and the modulation scheme employed is 4QAM. The ground station gain is configured at 40 dBi to enhance the transmission capabilities, while the satellite gain is set to 30 dBi. The transmission power is fixed at 100 W, and the noise temperature is maintained at 290 K \cite{NASA_GEO} to simulate typical thermal noise conditions in GEO. The jamming power introduced by the attacker is also set at 100 W. For spatial coverage, 5,000 different attacker positions are evaluated within the simulation, and 200 samples are generated for each configuration to ensure robust analysis. The radius of the VOI is specified as 5,000 km, capturing a critical area to assess the impact of jamming nearby. The simulation parameters used in this study are summarized in Table \ref{Table_sim}. In what follows, we present the performance results for stationary and time-variant jamming models.
 
The random forest model for the stationary jamming scenario, integrated with PCA, uses 100 estimators and a maximum depth of 10 to prevent overfitting. PCA addresses the challenges of correlated features, such as received power and SJNR, introduced by the attacker’s influence. By transforming these features into orthogonal components, PCA reduces multicollinearity and focuses on the most relevant jamming information. Additionally, PCA minimizes noise from feature variations, such as total received power and attacker positions, improving the model's robustness in environments with significant spatial variability.

\begin{table}[t!]
\renewcommand{\arraystretch}{1.2}
\caption{Simulation parameters}
\label{Table_sim}
\centering
\begin{tabular}{|l|l|}
    \hline
    \textbf{Parameter} & \textbf{Value} \\
    \hline
    Frequency (GHz) & 14 \\
    \hline
    Bandwidth (MHz) & 1 \\
    \hline
    Modulation & 4QAM \\
    \hline
    Ground station gain (dBi) & 40 \\
    \hline
    Satellite gain (dBi) & 30 \\
    \hline
    Transmission power (W) & 100 \\
    \hline
    Noise temperature (K) & 290 \\
    \hline
    Jamming power (W) & 100 \\
    \hline
    Number of attacker positions & 5000 \\
    \hline
   Number of samples & 200 \\
    \hline
    VOI radius (km) & 5000 \\
    \hline

\end{tabular}
\end{table}

\subsection{Stationary Model}
\label{sec:stationary_model}

\begin{table}[t!]
\centering
\caption{Performance evaluation of random forest model without PCA}
\begin{tabular}{|l|c|c|}
\hline
Class                   & Non-jammed  & Jammed \\ \hline
Training set size        & 2191           & 1809       \\ \hline
Testing set size         & 547            & 453        \\ \hline
Precision (\%)           & 70.0           & 71.0       \\ \hline
Recall (\%)              & 80.0           & 59.0       \\ \hline
F1 score (\%)            & 75.0           & 65.0       \\ \hline
Accuracy (\%)            & \multicolumn{2}{c|}{70.6}    \\ \hline
\end{tabular}
\label{table:performance_without_pca}
\end{table}

Fig. \ref{fig:combined_analysis} presents the feature selection and a comparative analysis of the confusion matrices with and without PCA. Subfigure (a) shows the correlation matrix of selected features, highlighting notable correlations such as the perfect correlation between the \(total\_received\_power\) and \(rss\) and the strong negative correlation between \(sjnr\_at\_target\) and \(is\_jammed\), which indicates that these features may significantly impact jamming detection accuracy. The selected features for the subsequent model training are $rss$, $distance\_to\_target$, $total\_received\_power$, $total\_amplitude\_mean$, $total\_amplitude\_std$, $total\_phase\_variance$. Subfigures (b) and (c) display the confusion matrices of the random forest model without and with PCA, respectively. The confusion matrices in (b) and (c) demonstrate the impact of PCA on the random forest model's performance for jamming detection. Without PCA (subfigure b), the model misclassifies 294 instances (110 false positives and 184 false negatives), achieving an accuracy of 70.6\% as shown in Table \ref{table:performance_without_pca}. With PCA applied and reduced to one dimension (Figure c), the misclassifications are significantly reduced to 70 instances (28 false positives and 42 false negatives), resulting in a much higher accuracy of 93.0\% as demonstrated in Table \ref{table:performance_with_pca}, indicating improved model efficiency with dimensionality reduction. This demonstrates that PCA effectively enhances model performance by focusing on the most informative features and minimizing noise.

\begin{table}[t!]
\centering
\caption{Performance evaluation of random forest model with PCA}
\begin{tabular}{|l|c|c|}
\hline
Class                   & Non-jammed  & Jammed \\ \hline
Training set size        & 2191           & 1809       \\ \hline
Testing set size         & 547            & 453        \\ \hline
Precision (\%)           & 93.0           & 94.0       \\ \hline
Recall (\%)              & 95.0           & 91.0       \\ \hline
F1 Score (\%)            & 94.0           & 92.0       \\ \hline
Accuracy (\%)            & \multicolumn{2}{c|}{93.0}    \\ \hline
\end{tabular}
\label{table:performance_with_pca}
\end{table}
\begin{figure}[!t]
  \centering
  \includegraphics[width=0.68\columnwidth]{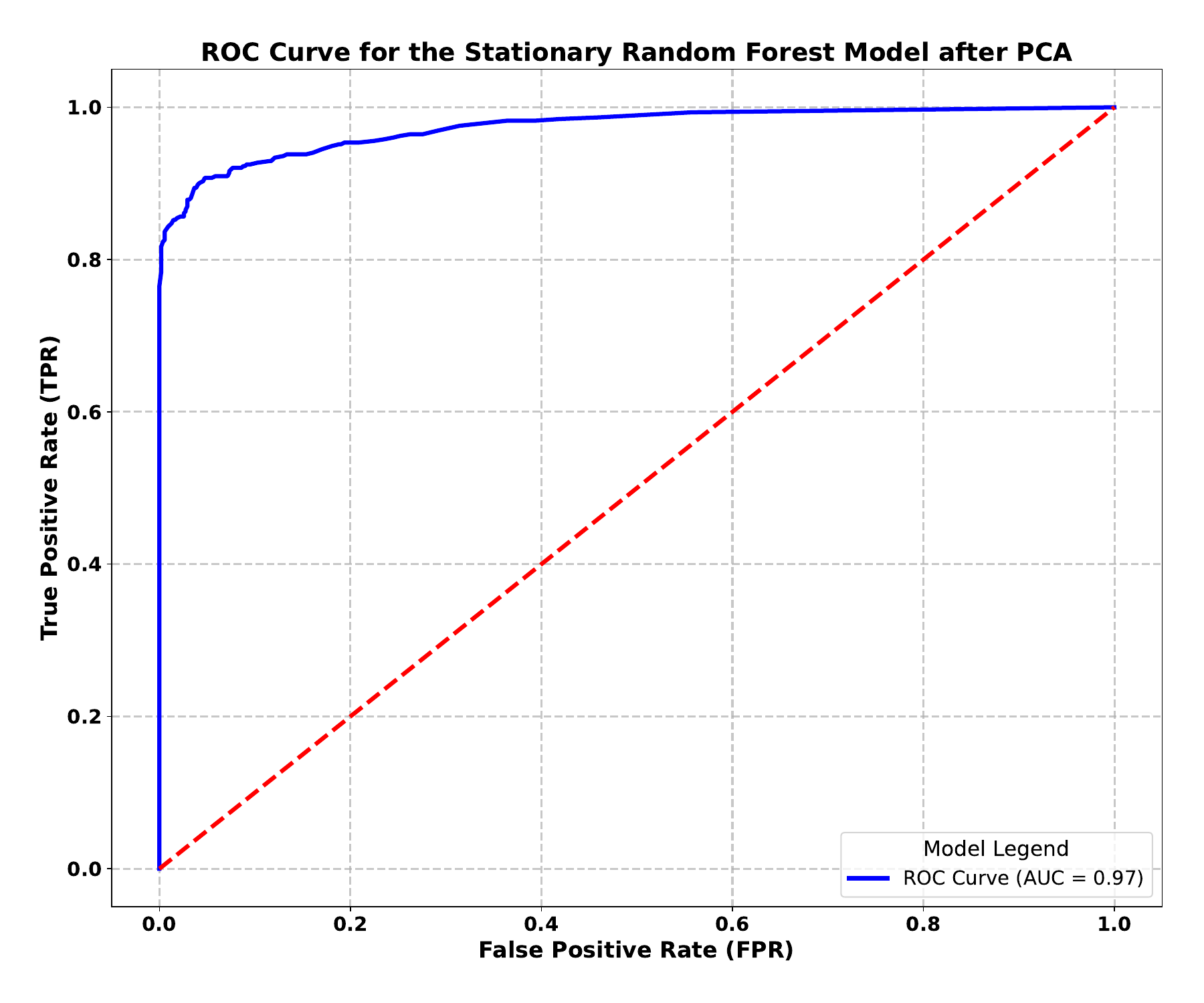}
  \caption{ROC curve for the random forest model in the stationary scenario after applying PCA}
  \label{fig:stationnary_ROC_curve}
\end{figure}

The receiver operating characteristic (ROC) in Fig. \ref{fig:stationnary_ROC_curve} demonstrates the performance of the random forest model in a stationary context after applying PCA. The blue line represents the model's true positive rate (TPR) against the false positive rate (FPR) across different threshold values. The ROC curve's high position near the top-left corner indicates strong classification performance, with an AUC (area under the curve) of 0.97, which signifies the excellent ability to distinguish between jammed and non-jammed states. The high AUC implies that applying PCA effectively improved the model’s classification accuracy by focusing on key features.
\subsection{Time-Variant Model}

In this section, we first explore the performance of the stationary-trained random forest model with PCA presented in subsection \ref{sec:stationary_model}. We applied this model to the time-variant data to predict jamming instances. However, as indicated by the performance metrics shown in Table \ref{table:performance_rf_pca}, the model significantly underperformed. The overall accuracy of 64.00\% ($\pm$ 19.00) and F1 Score of 66.00\% ($\pm$ 17.00), both with high standard deviations, suggest that the model consistently fails to perform well across different trajectories.

\begin{table}[t!]
\centering
\caption{Overall performance metrics across all trajectories for the stationary trained random forest model (with PCA)}
\begin{tabular}{|l|c|}
\hline
Metric          & Value ($\pm$ Standard Deviation) \\ \hline
Accuracy (\%)   & 64.00 $\pm$ 19.00 \\ \hline
F1 Score (\%)   & 66.00 $\pm$ 17.00 \\ \hline
\end{tabular}
\label{table:performance_rf_pca}
\end{table}

\begin{figure}[t!]
    \centering
    \begin{subfigure}[b]{0.38\textwidth}  
        \centering
        \includegraphics[width=\textwidth]{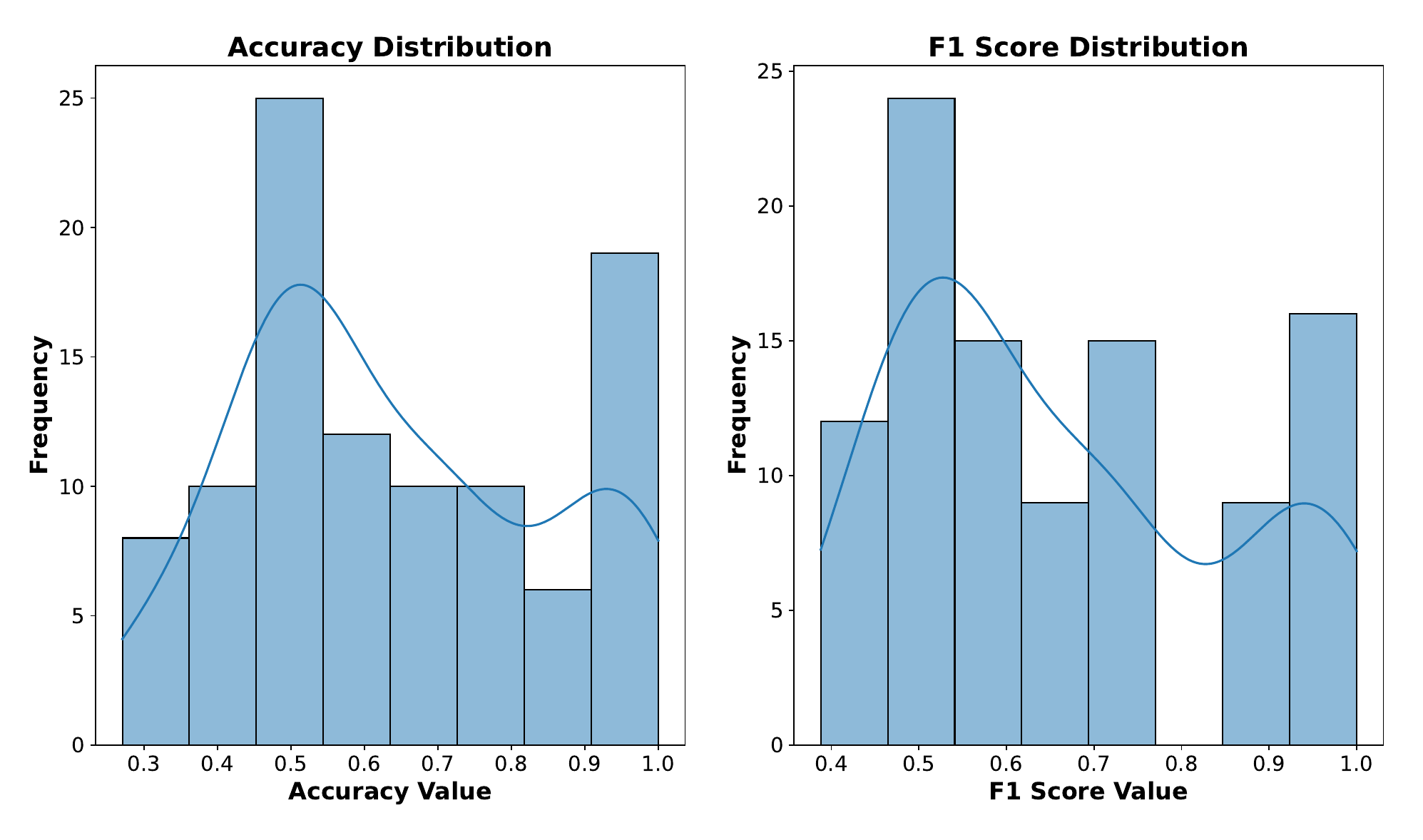}  
        \caption{Stationary trained random forest with PCA performance on the time-variant dataset}
        \label{fig:metrics_method1}
    \end{subfigure}
    
    \vspace{0.1cm}  

    \begin{subfigure}[b]{0.38\textwidth}  
        \centering
        \includegraphics[width=\textwidth]{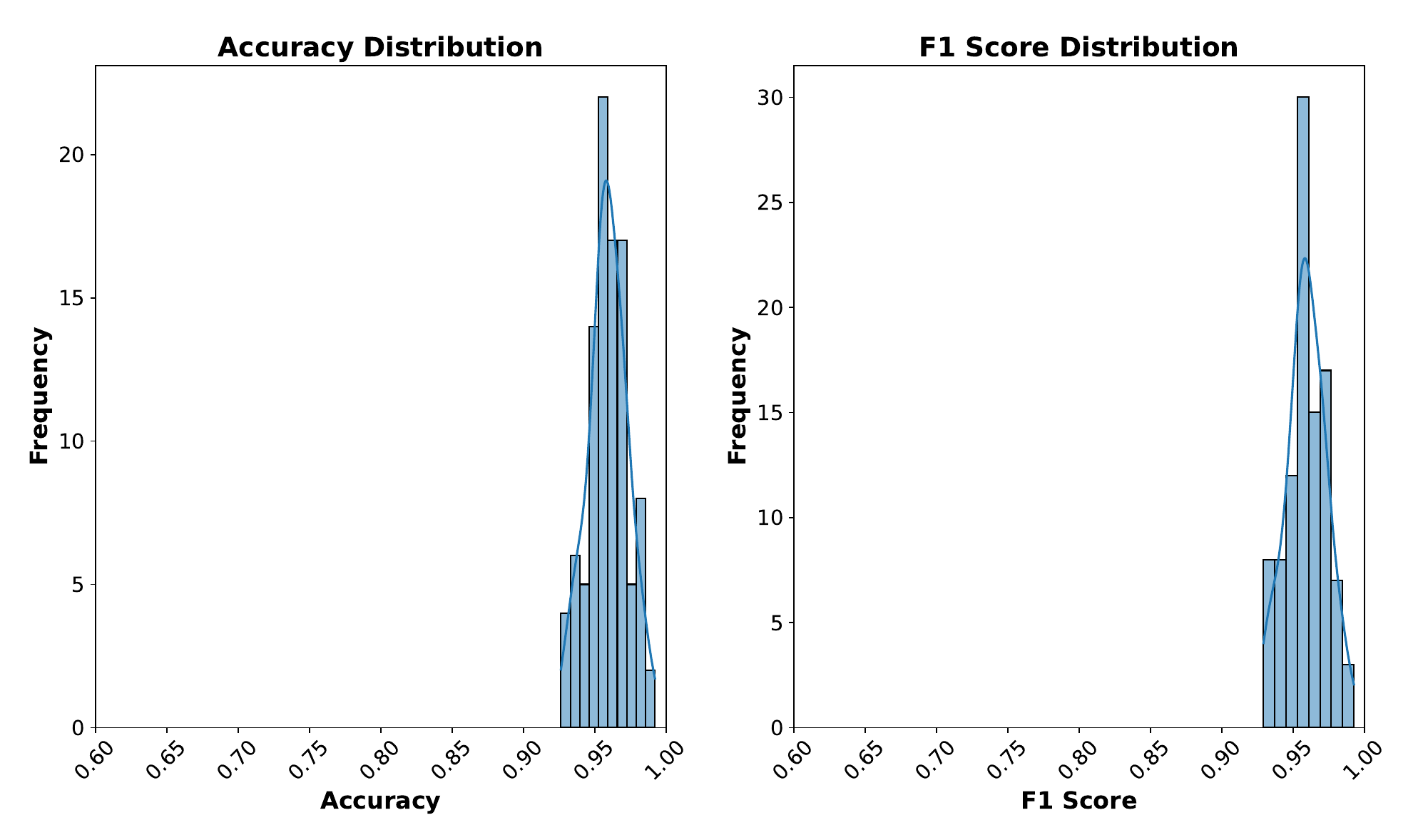}  
        \caption{Adaptive threshold technique performance on the time-variant dataset}
        \label{fig:metrics_method2}
    \end{subfigure}
    
    \caption{Time-variant comparison of performance metrics distribution for two methods over 100 generated trajectories on the time-variant dataset. (a) stationary trained random forest model with PCA (b) adaptive threshold technique}
    \label{fig:combined_metrics_comparison}
\end{figure}

\begin{figure}[t!]
  \centering
  \includegraphics[width=3.4in, height=2.2in]{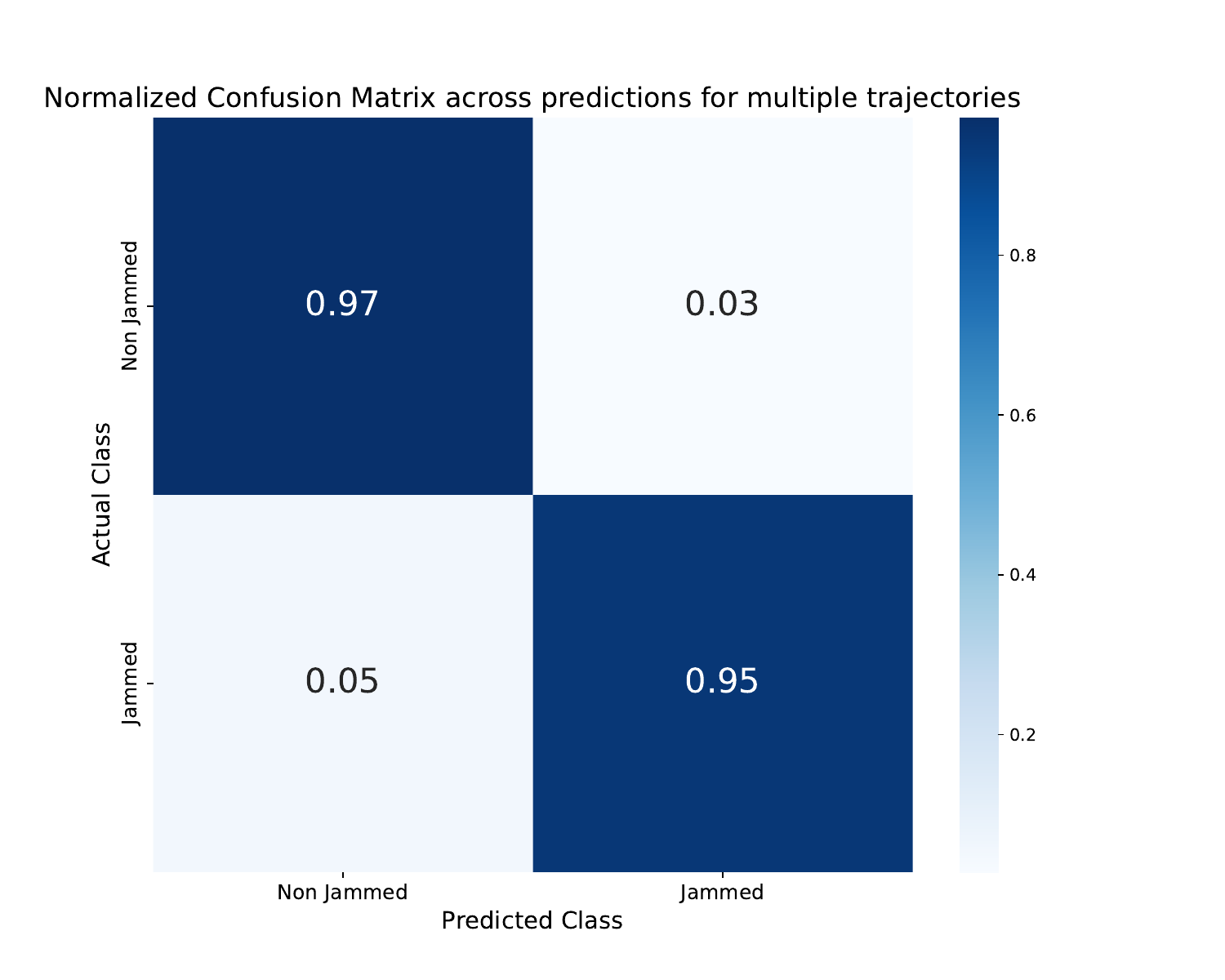}
  \caption{Confusion matrix for the adaptive thresholding approach in the time-variant model}
  \label{fig:confusion matrix time variant}
\end{figure}

\begin{table}[h!]
\centering
\caption{Performance evaluation of the adaptive threshold approach for the time-variant model}
\begin{tabular}{|l|c|c|}
\hline
Class                   & Non-jammed  & Jammed \\ \hline
Total prediction points & 25,890     & 26,394 \\ \hline
Precision (\%)           & 94.62          & 97.20      \\ \hline
Recall (\%)              & 97.23          & 94.57      \\ \hline
F1 Score (\%)            & 95.90          & 95.87      \\ \hline
Accuracy (\%)            & \multicolumn{2}{c|}{95.89}    \\ \hline
\end{tabular}
\label{table:performance_adaptive_threshold_time_variant}
\end{table}

A key issue with this model is the considerable variability in its metrics across different trajectories, as seen in Fig. \ref{fig:metrics_method1}. The large standard deviations show that the performance is highly inconsistent, which can be attributed to the different conditions present in each trajectory, such as the relative distances: the rate of change in the relative distance between the two satellites is influenced by the specific characteristics of the generated trajectories, resulting in varying dynamics of separation and approach, causing the model to perform poorly when these particular conditions arise. This inconsistency further emphasizes that the random forest model, trained on a stationary dataset, struggles to adapt to the dynamic changes in the time-variant conditions.

In contrast, in the time-variant model analysis, Fig. \ref{fig:confusion matrix time variant} presents a normalized confusion matrix, and Table \ref{table:performance_adaptive_threshold_time_variant} provides detailed performance metrics for the adaptive thresholding approach. The confusion matrix indicates high classification accuracy, with 97\% of non-jammed instances and 95\% of jammed instances correctly classified. Misclassification rates are low, with only 3\% of non-jammed and 5\% of jammed instances incorrectly classified. 

Table \ref{table:performance_adaptive_threshold_time_variant} further supports these results by showing high precision, recall, and F1 scores for both classes, with values above 94\% across all metrics. The overall accuracy of {95.89\%} reflects the model's robustness in adapting to time-variant conditions, effectively distinguishing between jammed and non-jammed states despite dynamic environmental changes. This high accuracy and balanced performance across metrics demonstrate that the adaptive thresholding approach performs well in the time-variant context.

\subsection{Deployment Strategy for Jamming Detection Algorithms}
Onboard deployment enables real-time detection and immediate response, which is critical in GEO to avoid latency from ground communication. It also enhances data security by reducing interception risks but is limited by computational constraints and the need for specialized hardware, increasing complexity.
In contrast, ground-based deployment allows for complex analysis using robust infrastructure and integrates data from multiple satellites for broader situational awareness. However, it introduces latency due to data transmission and exposes data to interception risks during downlink. To overcome the limitations of standalone onboard or ground-based deployments, a hybrid approach leverages the strengths of both: lightweight onboard algorithms provide real-time awareness on simple classification tasks while detailed analysis occurs on the ground. This balances responsiveness with a comprehensive evaluation to ensure effective jamming detection and mitigation.

\vspace{-2mm}
\section{Conclusion}
\label{Sec:concl}
In this paper, we introduced a novel scenario for space-based attacks in GEO, where an illegitimate GEO satellite performs active jamming on a {ground-to-satellite} communication link. We explored two jamming models: a stationary model, in which the attacker remains fixed, and a time-variant model, in which the attacker follows a designated trajectory over time. We developed two distinct detection methods to detect these threats: a random forest-based algorithm enhanced with PCA for improved accuracy in the stationary model and an adaptive thresholding approach tailored to the time-variant model. Our methodology emphasizes the integration of orbital dynamics, incorporating physical constraints from satellite dynamics to enhance detection accuracy and model robustness. Simulation results demonstrate the effectiveness of these methods, achieving high accuracy in identifying both static and dynamic jamming activities and highlighting the robustness of our detection strategies in addressing evolving jamming threats in GEO. 

Future work could expand upon the current solution by incorporating a more advanced physics-informed AI model, directly integrating physical principles into the detection algorithms. Further exploration of hybrid models that blend adaptive thresholding with machine learning could provide versatile detection capabilities across varied orbital scenarios. Finally, developing collaborative satellite networks for shared detection data and implementing proactive countermeasures are promising directions for strengthening GEO satellite link security against sophisticated on-orbit threats.

\bibliographystyle{IEEEtran}
\bibliography{reference}

\begin{thebibliography}{10}
\providecommand{\url}[1]{#1}
\csname url@samestyle\endcsname
\providecommand{\newblock}{\relax}
\providecommand{\bibinfo}[2]{#2}
\providecommand{\BIBentrySTDinterwordspacing}{\spaceskip=0pt\relax}
\providecommand{\BIBentryALTinterwordstretchfactor}{4}
\providecommand{\BIBentryALTinterwordspacing}{\spaceskip=\fontdimen2\font plus
\BIBentryALTinterwordstretchfactor\fontdimen3\font minus \fontdimen4\font\relax}
\providecommand{\BIBforeignlanguage}[2]{{%
\expandafter\ifx\csname l@#1\endcsname\relax
\typeout{** WARNING: IEEEtran.bst: No hyphenation pattern has been}%
\typeout{** loaded for the language `#1'. Using the pattern for}%
\typeout{** the default language instead.}%
\else
\language=\csname l@#1\endcsname
\fi
#2}}
\providecommand{\BIBdecl}{\relax}
\BIBdecl

\bibitem{jiang2015security}
C.~Jiang, X.~Wang, J.~Wang, H.-H. Chen, and Y.~Ren, ``Security in space information networks,'' \emph{IEEE Communications Magazine}, vol.~53, no.~8, pp. 82--88, 2015.

\bibitem{viasat-case-study}
{CyberPeace Institute}, ``{Case Study: Viasat},'' \url{https://cyberconflicts.cyberpeaceinstitute.org/law-and-policy/cases/viasat}, 2022, accessed: 2024-11-1.

\bibitem{zhang2024jamming}
Y.~Zhang, C.~Han, F.~Chu, W.~Xiong, and L.~Jia, ``Jamming analysis between non-cooperative mega-constellations based on satellite network capacity,'' \emph{Electronics}, vol.~13, no.~12, p. 2330, 2024.

\bibitem{10592290}
N.~Benchoubane, B.~Donmez, O.~Ben~Yahia, and G.~Karabulut~Kurt, ``Securing cislunar missions: A location-based authentication approach,'' in \emph{Security for Space Systems (3S)}, 2024, pp. 1--8.

\bibitem{han2021secure}
R.~Han, L.~Bai, C.~Jiang, J.~Liu, and J.~Choi, ``A secure architecture of relay-aided space information networks,'' \emph{IEEE Network}, vol.~35, no.~4, pp. 88--94, 2021.

\bibitem{lichtman2016analysis}
M.~Lichtman and J.~H. Reed, ``Analysis of reactive jamming against satellite communications,'' \emph{International Journal of Satellite Communications and Networking}, vol.~34, no.~2, pp. 195--210, 2016.

\bibitem{taricco2022jamming}
G.~Taricco and N.~Alagha, ``{On jamming detection methods for satellite Internet of Things networks},'' \emph{International Journal of Satellite Communications and Networking}, vol.~40, no.~3, pp. 177--190, 2022.

\bibitem{planta2024let}
U.~Planta, J.~Rederlechner, G.~Marra, and A.~Abbasi, ``Let me do it for you: On the feasibility of inter-satellite friendly jamming,'' in \emph{2024 Security for Space Systems (3S)}, 2024, pp. 1--6.

\bibitem{han2020dynamic}
C.~Han, A.~Liu, H.~Wang, L.~Huo, and X.~Liang, ``Dynamic anti-jamming coalition for satellite-enabled army {IoT: A} distributed game approach,'' \emph{IEEE Internet of Things Journal}, vol.~7, no.~11, pp. 10\,932--10\,944, 2020.

\bibitem{tang2024leo}
C.~Tang, J.~Ding, and L.~Zhang, ``{LEO} satellite downlink distributed jamming optimization method using a non-dominated sorting genetic algorithm,'' \emph{Remote Sensing}, vol.~16, no.~6, p. 1006, 2024.

\bibitem{Aiswarya2022Jamming}
A.~K and V.~S, ``Jamming attack detection using machine learning algorithms in wireless network,'' \emph{International Journal of Advanced Research in Science, Communication and Technology (IJARSCT)}, vol.~2, no.~1, pp. 1--8, August 2022.

\bibitem{9456965}
A.~Shafique, A.~Mehmood, and M.~Elhadef, ``Detecting signal spoofing attack in {UAV}s using machine learning models,'' \emph{IEEE Access}, vol.~9, pp. 93\,803--93\,815, 2021.

\bibitem{9666744}
G.~Aissou, H.~O. Slimane, S.~Benouadah, and N.~Kaabouch, ``Tree-based supervised machine learning models for detecting {GPS} spoofing attacks on {UAS},'' in \emph{Ubiquitous Computing, Electronics \& Mobile Communication Conference (UEMCON)}, 2021, pp. 0649--0653.

\bibitem{hofmann2018satellite}
C.~A. Hofmann and A.~Knopp, ``Satellite downlink jamming propagation measurements at ku-band,'' in \emph{IEEE Military Communications Conference (MILCOM)}, 2018, pp. 853--858.

\bibitem{threatsspace}
B.~Garino and J.~Gibson, ``Space system threats,'' in \emph{Space System Threats}.\hskip 1em plus 0.5em minus 0.4em\relax USAF, 2009, ch.~21, pp. 273--281.

\bibitem{NASA_GEO}
J.~Plante and B.~Lee, ``Environmental conditions for space flight hardware: {A} survey,'' https://ntrs.nasa.gov/citations/20060013394, 2005, accessed: 2024-11-13.

\end{thebibliography}
\end{document}